\documentclass[dvips]{article}
\usepackage{supertabular,lscape,epsfig}
\usepackage{amssymb}
\usepackage{amsmath}

\DeclareSymbolFont{ppa}{OT1}{ppl}{m}{it}
\DeclareMathSymbol{\vv}{\mathalpha}{ppa}{'166}

\begin{document}

\newcommand{\dd}{\,{\rm d}}
\newcommand{\ie}{{\it i.e.},\,}
\newcommand{\etal}{{\it et al.\ }}
\newcommand{\eg}{{\it e.g.},\,}
\newcommand{\cf}{{\it cf.\ }}
\newcommand{\vs}{{\it vs.\ }}
\newcommand{\zdot}{\makebox[0pt][l]{.}}
\newcommand{\up}[1]{\ifmmode^{\rm #1}\else$^{\rm #1}$\fi}
\newcommand{\dn}[1]{\ifmmode_{\rm #1}\else$_{\rm #1}$\fi}
\newcommand{\upd}{\up{d}}
\newcommand{\uph}{\up{h}}
\newcommand{\upm}{\up{m}}
\newcommand{\ups}{\up{s}}
\newcommand{\arcd}{\ifmmode^{\circ}\else$^{\circ}$\fi}
\newcommand{\arcm}{\ifmmode{'}\else$'$\fi}
\newcommand{\arcs}{\ifmmode{''}\else$''$\fi}
\newcommand{\MS}{{\rm M}\ifmmode_{\odot}\else$_{\odot}$\fi}
\newcommand{\RS}{{\rm R}\ifmmode_{\odot}\else$_{\odot}$\fi}
\newcommand{\LS}{{\rm L}\ifmmode_{\odot}\else$_{\odot}$\fi}

\newcommand{\Abstract}[2]{{\footnotesize\begin{center}ABSTRACT\end{center}
\vspace{1mm}\par#1\par
\noindent
{~}{\it #2}}}

\newcommand{\TabCap}[2]{\begin{center}\parbox[t]{#1}{\begin{center}
  \small {\spaceskip 2pt plus 1pt minus 1pt T a b l e}
  \refstepcounter{table}\thetable \\[2mm]
  \footnotesize #2 \end{center}}\end{center}}

\newcommand{\TableSep}[2]{\begin{table}[p]\vspace{#1}
\TabCap{#2}\end{table}}

\newcommand{\FigCap}[1]{\footnotesize\par\noindent Fig.\  %
  \refstepcounter{figure}\thefigure. #1\par}

\newcommand{\TableFont}{\footnotesize}
\newcommand{\TableFontIt}{\ttit}
\newcommand{\SetTableFont}[1]{\renewcommand{\TableFont}{#1}}

\newcommand{\MakeTable}[4]{\begin{table}[htb]\TabCap{#2}{#3}
  \begin{center} \TableFont \begin{tabular}{#1} #4 
  \end{tabular}\end{center}\end{table}}

\newcommand{\MakeTableSep}[4]{\begin{table}[p]\TabCap{#2}{#3}
  \begin{center} \TableFont \begin{tabular}{#1} #4 
  \end{tabular}\end{center}\end{table}}

\newenvironment{references}%
{
\footnotesize \frenchspacing
\renewcommand{\thesection}{}
\renewcommand{\in}{{\rm in }}
\renewcommand{\AA}{Astron.\ Astrophys.}
\newcommand{\AAS}{Astron.~Astrophys.~Suppl.~Ser.}
\newcommand{\ApJ}{Astrophys.\ J.}
\newcommand{\ApJS}{Astrophys.\ J.~Suppl.~Ser.}
\newcommand{\ApJL}{Astrophys.\ J.~Letters}
\newcommand{\AJ}{Astron.\ J.}
\newcommand{\IBVS}{IBVS}
\newcommand{\PASP}{P.A.S.P.}
\newcommand{\Acta}{Acta Astron.}
\newcommand{\MNRAS}{MNRAS}
\renewcommand{\and}{{\rm and }}
\section{{\rm REFERENCES}}
\sloppy \hyphenpenalty10000
\begin{list}{}{\leftmargin1cm\listparindent-1cm
\itemindent\listparindent\parsep0pt\itemsep0pt}}%
{\end{list}\vspace{2mm}}

\def\TYLDA{~}
\newlength{\DW}
\settowidth{\DW}{0}
\newcommand{\dw}{\hspace{\DW}}

\newcommand{\refitem}[5]{\item[]{#1} #2%
\def\REFARG{#3}\ifx\REFARG\TYLDA\else, {\it#3}\fi
\def\REFARG{#4}\ifx\REFARG\TYLDA\else, {\bf#4}\fi
\def\REFARG{#5}\ifx\REFARG\TYLDA\else, {#5}\fi.}

\newcommand{\Section}[1]{\section{#1}}
\newcommand{\Subsection}[1]{\subsection{#1}}
\newcommand{\Acknow}[1]{\par\vspace{5mm}{\bf Acknowledgements.} #1}
\pagestyle{myheadings}

\newcommand{\uprule}{\rule{0pt}{2.5ex}}
\newcommand{\douprule}{\rule[-2ex]{0pt}{4.5ex}}
\newcommand{\dorule}{\rule[-2ex]{0pt}{2ex}}

\newfont{\bb}{ptmbi8t at 12pt}
\newcommand{\xrule}{\rule{0pt}{2.5ex}}
\newcommand{\xxrule}{\rule[-1.8ex]{0pt}{4.5ex}}
\def\thefootnote{\fnsymbol{footnote}}
\begin{center}
{\Large\bf The Optical Gravitational Lensing Experiment.\\
\vskip3pt
Is Interstellar Extinction Toward the Galactic \\
\vskip3pt
Center Anomalous?\footnote{Based on  observations obtained with the 1.3~m
Warsaw telescope at the Las Campanas  Observatory of the Carnegie
Institution of Washington.}}

\vskip1cm
{\bf A.~~U~d~a~l~s~k~i}
\vskip3mm
{Warsaw University Observatory, Al.~Ujazdowskie~4, 00-478~Warszawa,
Poland\\
e-mail: udalski@astrouw.edu.pl}
\end{center}

\Abstract{Photometry of the Galactic bulge, collected during the OGLE-II
microlensing search, indicates high and non-uniform interstellar
extinction toward the observed fields. We use the mean {\it I}-band
magnitude and $V-I$ color of red clump stars as a tracer of interstellar
extinction toward four small regions of the Galactic bulge with highly
variable reddening. Similar test is performed for the most reddened
region observed in the LMC.

We find that the slope of the location of red clump stars in the
color-magnitude diagrams (CMDs) in the Galactic bulge is significantly
smaller than the slope of the reddening line following the standard
extinction law (${\rm R}_V=3.1$) for approximations of the extinction
curve by both Cardelli, Clayton  and Mathis (1989, CCM89) and
Fitzpatrick (1999, F99). The differences are much larger for the CCM89
approximation which, on the other hand, indicates the same slopes for
the control field in the LMC, contrary to the F99 approximation.

We discuss possible systematic effects that could cause the observed
discrepancy. Anomalous extinction toward the Galactic bulge seems to be
the most natural explanation. Our data indicate that, generally,  the
ratio of the total to selective absorption, ${\rm R}_{\it VI}$, is much
smaller toward the Galactic bulge than the value corresponding to the
standard extinction curve (${\rm R}_V=3.1$). However, ${\rm R}_{\it VI}$
varies from one line-of-sight to another. 

Our results explain why the red clump and RR~Lyr stars in the Baade's
window dereddened with standard value of ${\rm R}_{\it VI}$  are redder
compared to those of the local population.
} 

\Section{Introduction} 

Optical photometric observations of stellar populations in the Galactic
center (GC) regions are, for many reasons, very difficult. First of all
these regions are highly obscured by interstellar extinction. On the
other hand, the stellar density in the so called windows of lower
extinction  is usually so high that good resolution and precise
photometry can only be obtained from very good astronomical sites with
modern CCD detectors. It is not surprising then that very limited
photometric coverage of the Galactic bulge (GB) can be found in the
literature. The existing papers concentrated typically on the well known
Baade's window or selected objects in the bulge, \eg globular clusters.
The situation has significantly changed in 1990s when large microlensing
surveys like MACHO, EROS or OGLE started using the GB stars as sources
for microlensing. Huge photometric databases covering large areas of the
GB were created as a natural by-product of the microlensing searches.
For example, the OGLE project has recently released the photometric maps
of the GB  containing photometry of about 30 million stars from 49
fields covering 11 square degrees in different regions of the GB
(Udalski \etal 2002). These data can be very useful for many
astrophysical projects including studying the stellar populations in
many lines-of-sight in the GB.

Good understanding of properties of interstellar extinction  is crucial
for studying the GB. Unfortunately, it is large and highly non-uniform
over the bulge making dereddening of observed magnitudes and colors
difficult and uncertain. One of the possible solutions is to observe the
GB stars in the infrared where the interstellar extinction is much
smaller. Indeed, large areas of the GB were covered by the large field
infrared surveys -- DENIS and 2MASS. However, the range of both surveys
was so shallow that only the brightest objects from the GB could be
reliably measured. Nevertheless, infrared data from both surveys were
used for construction of {\it K}-band extinction maps in the GB  by
Schultheis \etal (1999) and Dutra \etal (2002ab).

The optical range still remains very important for studying the GB. In
this range the ({\it V},$V-I$) extinction maps were compiled by Stanek
(1996). They cover the Baade's window and were based on the method of
Wo{\'z}niak and Stanek (1996) who used the mean brightness and color
of red clump giants as an indicator of the amount of interstellar
extinction. After corrections of the zero point (Gould \etal 1998 and
Alcock \etal 1998) these maps have been widely used for many GB
projects.

Dereddening of the GB red clump and RR~Lyr stars with Stanek's maps led,
however, to somewhat surprising results. Paczy{\'n}ski and Stanek (1998)
found that the mean dereddened $V-I$ color of the GB red clump stars is
redder by about 0.2~mag  than the color of nearby red clump giants. As
the metallicity of both populations is similar (Udalski 2000, Ramirez
\etal 2000), large difference in dereddened colors, decreased later to
0.11~mag by Paczy{\'n}ski \etal (1999) based on more accurate OGLE-II
photometry, made this result suspicious. Moreover, similar effect was
found for the GB RR~Lyr stars observed by OGLE and dereddened with
Stanek's maps (Stutz, Popowski and Gould 1999). Popowski (2000)
summarized possible explanations of these discrepancies. He noted that
while it is in principle possible that the differences are caused by
different properties of the local and bulge populations of red giants or
RR Lyr stars as proposed by Paczy{\'n}ski (1998) and Stutz \etal (1999),
the simplest and most plausible explanation would be  non-standard
properties of interstellar extinction. The discrepancy would vanish if
the ratio of the total to selective absorption, ${\rm R}_{\it VI}={\rm
A}_{\it V}/{\rm E}(V-I)$, was equal to about 2.1 instead of the standard
value of about 2.5. Thus, he suggested for the first time that the
interstellar extinction toward the GB might be anomalous what can
significantly affect any analysis of the GB properties. 

The problem becomes more important and urgent because of rapidly
increasing number of the GB observations with new generation of large
telescopes.  Gould \etal (2001) reviewed existing results of
determination of ${\rm R}_{\it VI}$ and proposed a new method. It was
based on comparison of {\it VIK} magnitudes of red giants with
magnitudes of the local population of similar stars. Unfortunately,
their results were inconclusive, mostly due to uncertainties of zero
points of the GB photometry. The question whether the properties of
interstellar extinction toward the GB are anomalous or close to the
standard ones still remains open.

In this paper we present results of analysis of photometric data of
several GB fields from the OGLE-II microlensing survey (Udalski \etal
2002) with the interstellar extinction high and highly variable across
the field. Straightforward analysis suggests that ${\rm R}_{\it VI}$ in
the GB is different than the value corresponding to the standard
extinction curve (${\rm R}_V=3.1$). In general it is much smaller but
varies along different lines-of-sight.

\Section{Observational Data}

The photometric data used in this paper come from the recently released
``OGLE-II {\it VI} Maps of the Galactic Bulge'' (Udalski \etal
2002)\footnote{ {\it http://www.astrouw.edu.pl/\~{}ogle} or {\it
http://bulge.princeton.edu/\~{}ogle}}. The maps contain {\it VI}
photometry and astrometry of about 30 million stars from the GB.
Provided photometry is the mean photometry from typically a few hundreds
measurements in the {\it I}-band and several measurements in the {\it
V}-band collected in 1997--2000. Accuracy of the zero points of
photometry is about 0.04~mag. Quality of data and possible systematic
errors are discussed in detail in Udalski \etal (2002).

The observed fields, $14.2\times57$ arcmin each,  are located in many
lines-of-sight in the GB (see the map -- Fig.~1 in Udalski \etal 2002).
Several of them are located in the regions where the interstellar
extinction is high or highly variable across the field. Udalski \etal
(2002) present several $({\it I},V-I)$ color-magnitude diagrams (CMDs)
of such fields. They show in a spectacular way how the interstellar
extinction reddens the GB stars when observing closer and closer to the
GC. In many fields where extinction is patchy or varies significantly
across the field, the compact and round feature of the CMD like the red
clump becomes not only highly elongated but rather a strip of stars
distributed along the reddening line. Large range of these variations
enables us to  study directly the properties of interstellar extinction
toward the GB.

For further analysis we selected the following regions from the OGLE-II
maps: A) the region closest to the GC consisting of fields BUL$\_$SC5,
BUL$\_$SC37, BUL$\_$SC3, BUL$\_$SC4 and BUL$\_$SC39. Centers of these
fields are located from $1\zdot\arcd3$ to $2\zdot\arcd3$ from the
Galactic center. B) the region of patchy extinction located about
$3\zdot\arcd2$ from the GC: BUL$\_$SC22 and BUL$\_$SC23. C) BUL$\_$SC43:
a field at the opposite side of the GB at positive latitude, located
about $3\arcd$ from the GC. D) Baade's window fields: BUL$\_$SC46,
BUL$\_$SC45, BUL$\_$SC1 and BUL$\_$SC38. E) Comparison field from the
LMC including two LMC fields with the largest variation of interstellar
extinction: LMC$\_$SC16 and LMC$\_$SC17. The data for this region come
from similar OGLE-II maps of the LMC (Udalski \etal 2000).

\Section{Data Analysis}

To trace the amount of interstellar extinction toward the GB we selected
the GB red clump (RC) giants as a reference source of light similar to
Wo{\'z}niak and Stanek (1996) and Stanek (1996). The RC stars constitute
a compact and well defined group in the CMD and were proposed as a good
and well calibrated distance indicator (Paczy{\'n}ski and Stanek 1998).
Although there is no common agreement on the magnitude of population
effects (age, metallicity) on the mean magnitude of RC stars 
(Pietrzy{\'n}ski, Gieren and Udalski 2003, Girardi and Salaris 2002), it
is rather widely accepted that within a population it is constant to a
few hundredths of a magnitude. As the selected regions are small -- at
most 1.3 square degree for the region (A) -- we assume that the mean
properties of the RC stars are similar within a field and the RC mean
magnitude and color are constant across the region.

Contrary to Wo{\'z}niak and Stanek (1996) and Stanek (1996) we analyzed
the mean RC magnitude in the $({\it I}, V-I)$ CMD instead of $({\it V},
V-I)$ as they did. Paczy{\'n}ski and Stanek (1998) first showed that the
{\it I}-band magnitude of the RC is constant with color. The {\it
V}-band magnitude is a function of $V-I$ what can lead to systematic
errors when analyzing interstellar extinction effects (Popowski 2000).

We divided each of the analyzed OGLE-II fields into 64 subfields of
$512\times 512$ pixel size ($3\zdot\arcm5\times $3\zdot\arcm5 on the
sky). $({\it I}, V-I)$ CMD of each of the subfields was retrieved from
the OGLE-II maps, and centroid of the RC was determined.  In spite of so
fine grid, in some of the subfields (especially those located close to
the GC) the variations of interstellar extinction even over so small
area were so significant that the RC remained severely elongated. Such
fields were excluded from further analysis. The typical accuracy of the
RC centroids is about 0.04 mag in {\it I} and $V-I$.

\begin{figure}[htb]

\centerline{\includegraphics[width=12.5cm, bb=35 45 515 420]{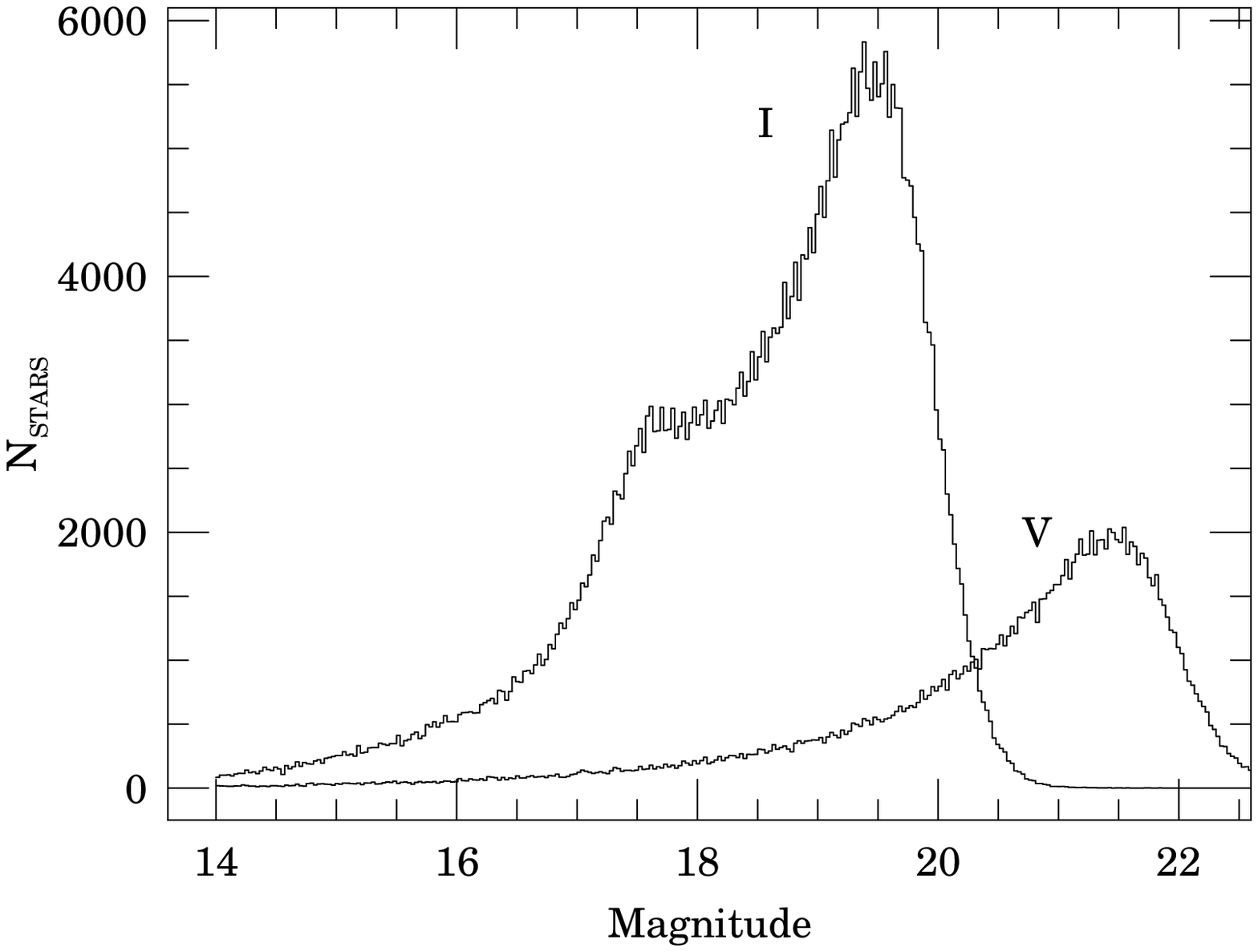}}
\FigCap{Distribution of {\it V} and {\it I}-band magnitudes of stars in
the BUL$\_$SC5 field (about 150000 and 450000 stars for {\it V} and {\it
I}, respectively). The bins are 0.03~mag wide.}
\end{figure}

As the interstellar extinction increases, the {\it V} and {\it I}-band
magnitudes of the RC become fainter and for significantly reddened
fields they can reach the photometry  limit making the determined mean
magnitudes biased and unreliable. In practice, the problem affects only
our most reddened field BUL$\_$SC5, located closest to the Galactic
center. To avoid the problem we limited our subfields to those which do
not suffer from incompleteness of photometry. Fig.~1 presents the
distribution of stars in the field BUL$\_$SC5 as a function of
magnitude. From the shape of the histogram we may conclude that the
OGLE-II photometry is complete up to ${\it V}\approx 21.3$~mag and ${\it
I}\approx 19.4$~mag. In the {\it I}-band the brightness of RC stars is
always far from the incompleteness limit. On the other hand, in the {\it
V}-band it reaches the limit for many highly reddened subfields.
Therefore, we set a safety margin of 0.6~mag and removed from further
analysis all subframes with the mean magnitude of RC ${\it V}>20.7$~mag.
The margin is large enough to have the mean magnitude of the RC reliably
measured. In other fields the completeness of photometry starts to drop
at slightly (0.2--0.3~mag) brighter magnitudes due to higher crowding
but, on the other hand, those fields are less reddened than BUL$\_$SC5
field and the RC brightness is always far from the incompleteness limit.
Also the brightness of the RC in the LMC is far away from the
incompleteness limit of the LMC photometry.

Assuming that the variation of the mean RC magnitude and color are due
to interstellar extinction,  the ${\rm R}_{\it VI}$ ratio can be
directly determined by fitting a straight line to the location of RC
centroids in the $({\it I}, V-I)$ CMD (plus 1.0). Figs.~2--5 present
CMDs with location of RC centroids (asterisks) for regions (A)--(D),
respectively. Solid line is a least square fit to the data. Slopes of
the fitted lines ($\Delta I/\Delta (V-I)$) are listed in the second
column of Table~1. The RC centroids and the fit are plotted twice in
Figs.~2--5 what will be discussed in the next Section. The magnitude
scale refers to the lower plot, the upper one is shifted by 1.2~mag up
along the magnitude axis.

\begin{figure}[p]

\centerline{\includegraphics[width=12.5cm, bb=40 60 520 310]{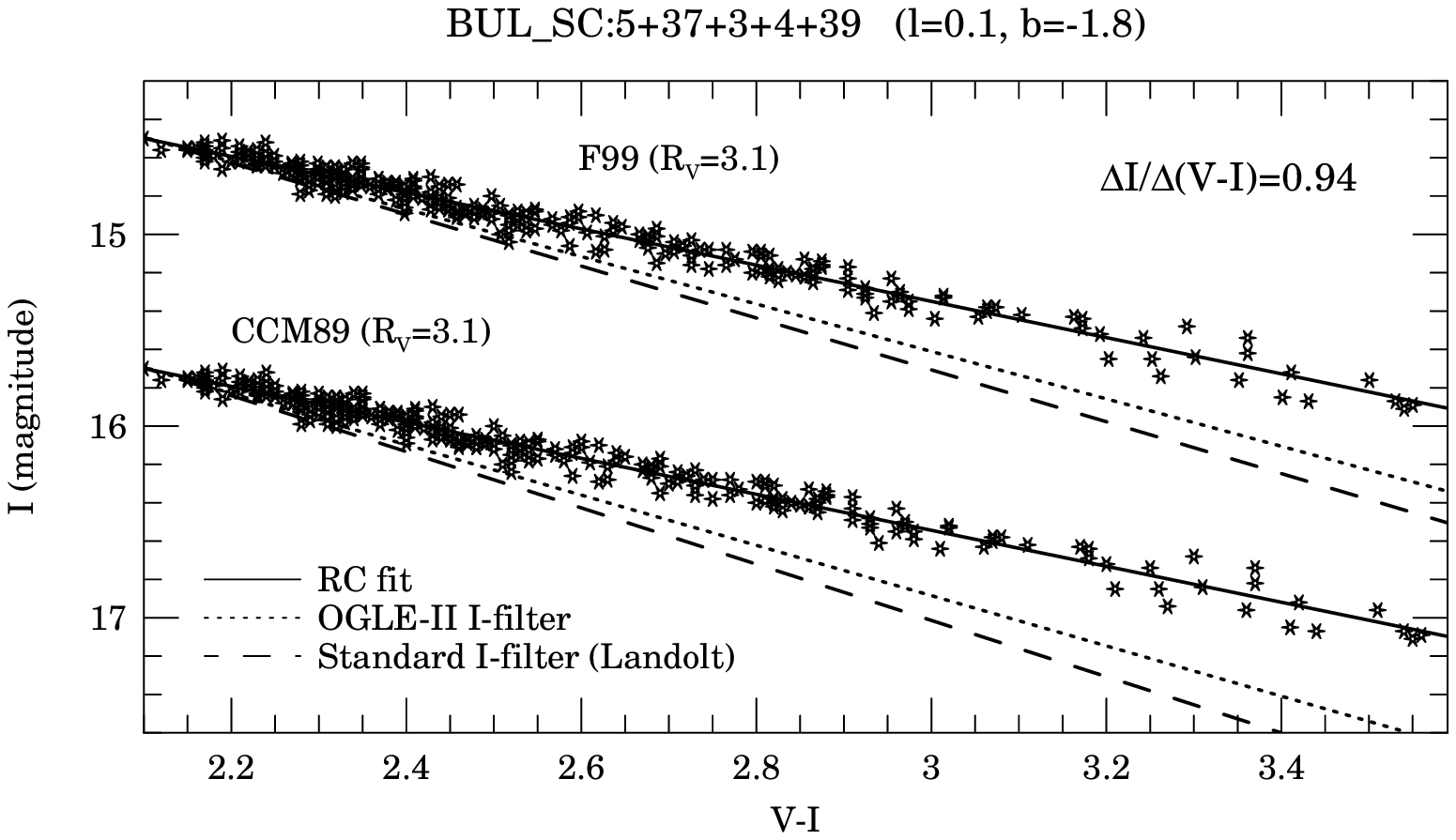}}
\vskip-5pt

\FigCap{Location of centroids of RC stars (asterisks) in the {\it I} \vs
$V-I$ CMD in the region (A). Solid line marks the best linear fit which
slope, $\Delta I/\Delta (V-I)$, is provided in the upper right corner.
Dashed line indicates reddening line for the standard extinction (${\rm
R}_V=3.1$) seen through the standard {\it I}-band filter, while the
dotted line the standard reddening line through the OGLE-II {\it I}
filter. Lower and upper plots correspond to CCM89 and F99 approximations
of the extinction curve, respectively (see text for details). The
magnitude scale refers to the lower plot; the upper plot is shifted up
by 1.2~mag.}
\vskip1.5cm
\centerline{\includegraphics[width=12.5cm, bb=40 60 520 310]{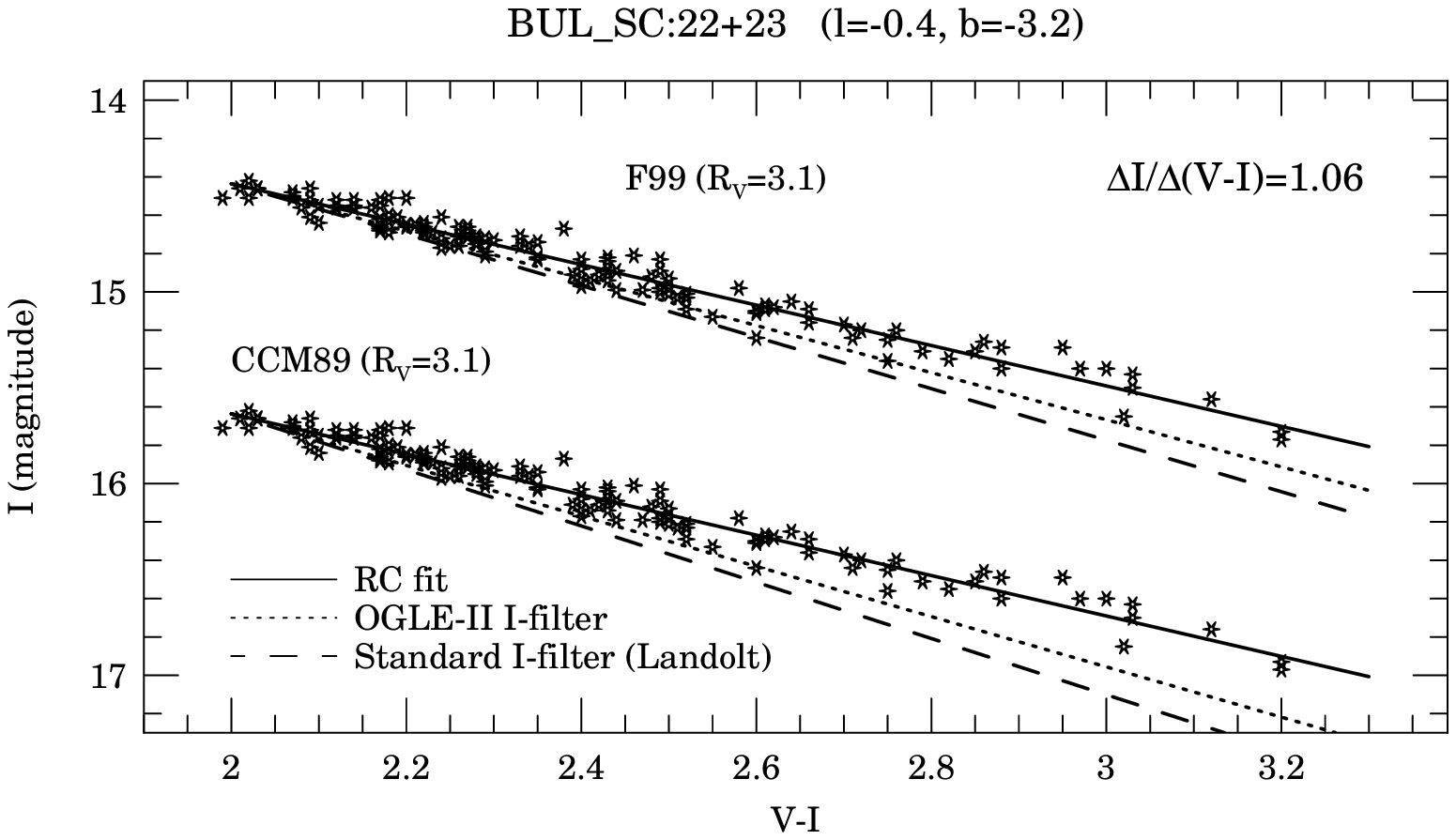}}
\vskip-5pt
\FigCap{Same as in Fig.~1, for region (B).}
\end{figure}

\begin{figure}[p]
\centerline{\includegraphics[width=12.5cm, bb=40 60 520 310]{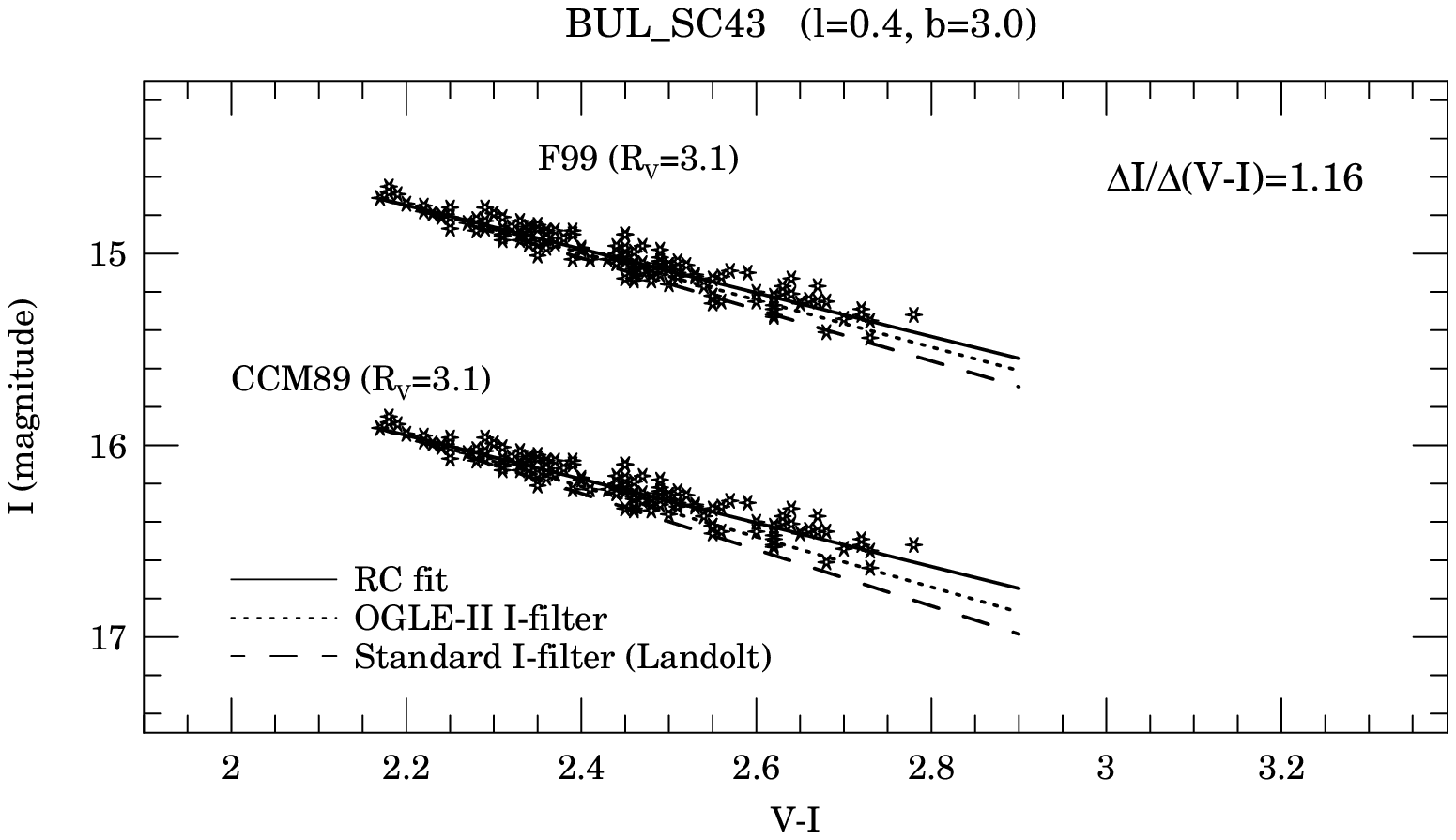}}
\vskip-5pt
\FigCap{Same as in Fig.~1, for region (C).}
\vskip1.5cm
\centerline{\includegraphics[width=12.5cm, bb=40 60 520 310]{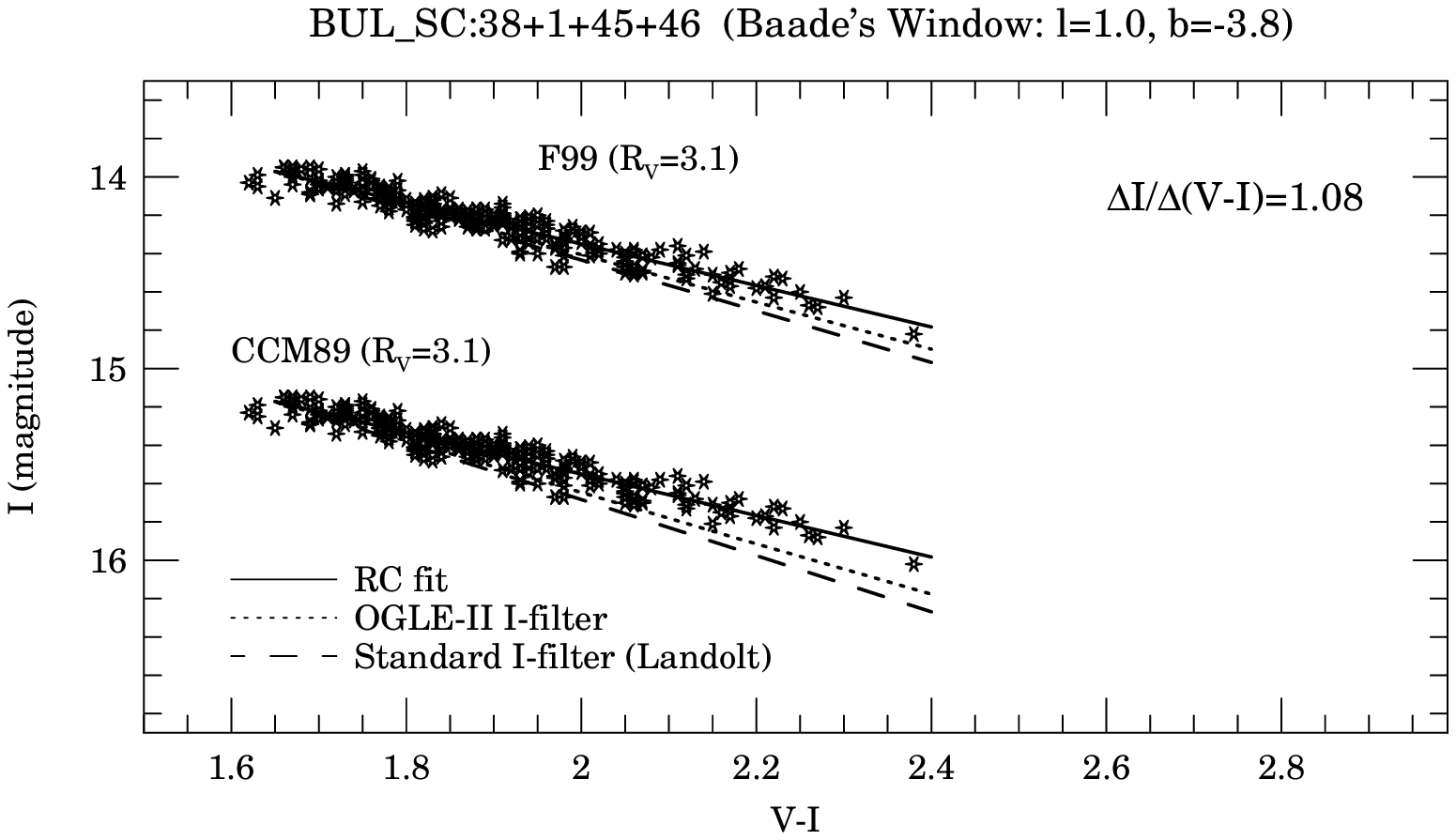}}
\vskip-5pt
\FigCap{Same as in Fig.~1, for region (D).}
\end{figure}

\Section{Discussion}

Figs.~2--5 indicate that the variation of interstellar extinction in the
selected regions is indeed very large. In particular, in  the region (A)
the range of observed $V-I$ colors of RC stars reaches almost 1.5~mag,
in region (B) --  1.2~mag. Thus, the base for studying the interstellar
extinction properties is very extended. It should be also noted that
because the photometric data come from the same instrument the results
are insensitive to some systematic effects like uncertainty of zero
points etc.

\begin{figure}[t]
\centerline{\includegraphics[width=12.5cm, bb=40 60 520 310]{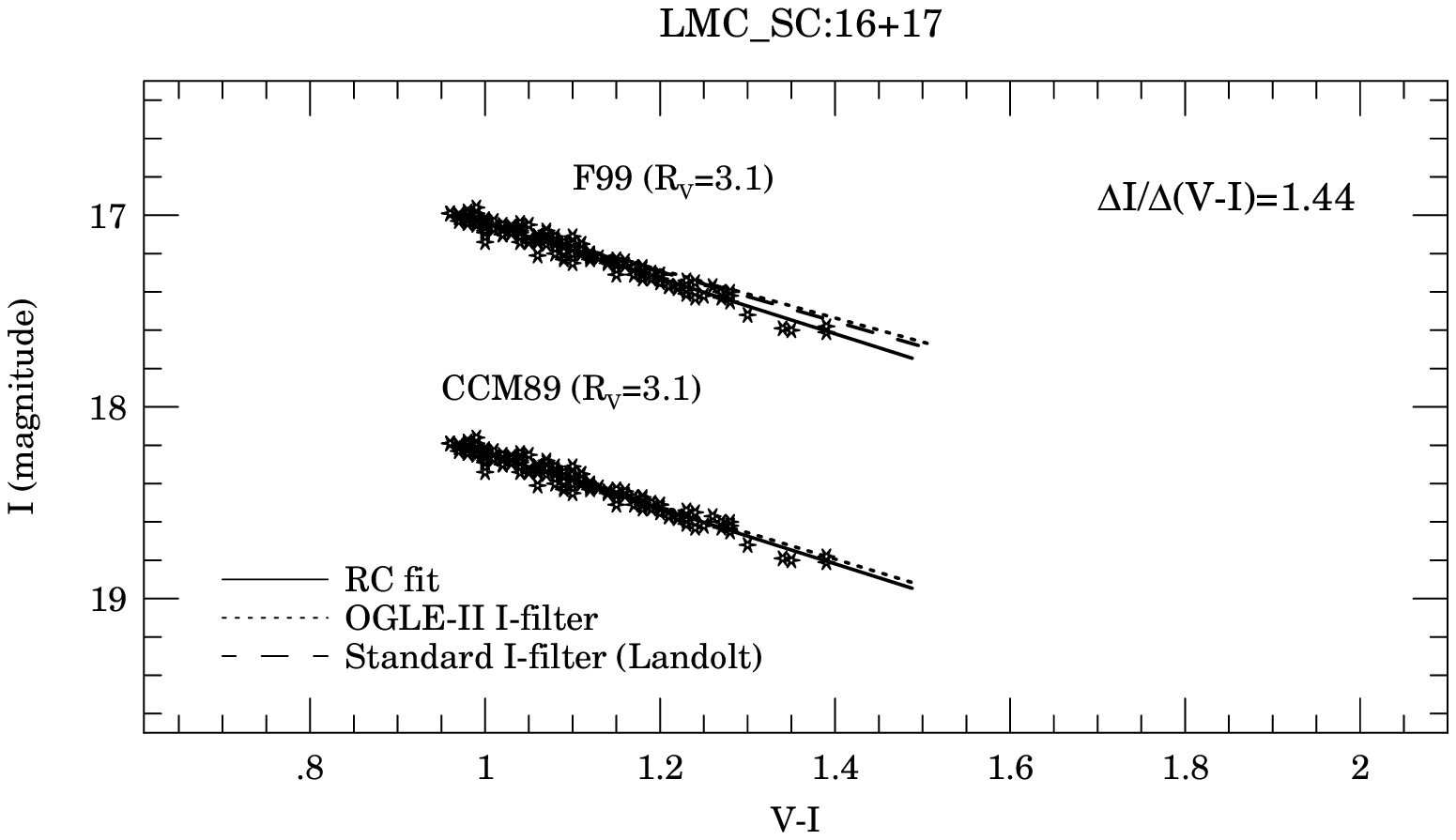}}
\vskip-5pt
\FigCap{Same as in Fig.~1, for region (E). Dashed line overlaps with the
solid line in the lower (CCM89) plot.}
\end{figure}

It is striking that the slopes of the best linear fit to the RC
centroids listed in Table~1 are significantly smaller than the slope of
the ({\it I},$V-I$) standard extinction reddening line (usually assumed
to be  about 1.5). This may indicate that the interstellar extinction
toward the Galactic bulge is indeed anomalous, and the ${\rm R}_{\it
VI}$ ratio is significantly smaller than the standard value, as
predicted by Popowski (2000). To check this possibility we analyzed
possible systematic effects which could lead to smaller slope of the RC
centroids as observed in the OGLE-II data.

First, we repeated the procedure described in the previous section for
the comparison field (E) located in the LMC. This region covers the area
of the largest variation of extinction  in the observed OGLE-II fields
in the LMC. Although the range of color changes is smaller in this case
than in the GB regions, centroids of the RC can be derived much more
accurately, because the RC is much more compact. Fig.~6 shows the RC
centroids and the fit similarly to  the GB fields. The slope of the best
linear fit to the LMC data is also provided in Table~1. One can
immediately conclude that in the LMC case the properties of interstellar
extinction and ${\rm R}_{\it VI}$ ratio in this direction are similar to
the standard ones. This strengthens our suspicion that the extinction
toward the GB might be anomalous.

Then, we considered possible instrumental effects. As noted by Udalski
\etal (2002) the {\it I}-band filter used by OGLE-II was an often used
approximation of the {\it I}-band with a Schott glass filter. However,
its transmission curve has somewhat wider red-wing what may lead to
systematic deviations from the standard values for very red stars.
Udalski \etal (2002) presented detailed modeling of the possible
discrepancies between the OGLE-II and standard {\it I}-band filter using
the Kurucz's model atmosphere of a typical GB red clump star, reddened
with the standard interstellar extinction (CCM89) similar as in the
observed GB fields. The differences noted in Udalski \etal (2002) work
in the direction of lowering the slope of the reddening line in the CMD
diagram.  Because the regions analyzed in this paper are in the color
range where those differences become significant, the non-matching
filters could, in principle, introduce a systematic effect leading to
the observed discrepancy of slopes. On the other hand the differences
between the OGLE-II {\it I}-band filter and the standard one, as modeled
by Udalski \etal (2002), are much too small to explain so different
slopes of RC stars as observed in Figs.~2--5. Different slope in the
region (D) where the large part of RC magnitudes is in the range of
photometry well calibrated by standard stars seems to indicate that the
effect is not of instrumental origin.

Ideally, one should either calibrate the red range of the  OGLE-II
photometry with standards or compare with other photometry of highly
reddened fields taken with the standard {\it I}-band filter.
Unfortunately, none of these options is feasible. There is practically
no standards with $V-I>2$ as those very red stars are usually variable.
Also, as mentioned in Introduction, as far as we know there is no
precise CCD photometry of highly reddened GB fields in the literature to
which we could directly compare our RC photometry. We hope that our
paper will encourage astronomers to carry out such project. It is worth
noting that highly reddened RC stars could be ideal standard stars for
calibrating very red photometry, as they are usually non-variable,
bright enough and, due to high interstellar extinction, could be found
in relatively uncrowded regions.

The only available photometry of required accuracy we could find in the
literature for comparison with our data comes from the Baade's window.
Stanek \etal (2000) observed a small $9'\times 9'$ field of relatively
small interstellar extinction while Richtler \etal (1998) even smaller
field around the globular cluster NGC~6528. Photometric data from both
projects are available {\it via} Internet so we could compare
photometries on the star by star basis. Stanek \etal (2000) and Richtler
\etal (1998) fields are located in our BUL$\_$SC45 and BUL$\_$SC46
fields, respectively. None of them contains significantly reddened RC
stars but, fortunately, they contain many red stars. On the other hand,
red stars in the GB are usually variable (Mizerski and Bejger 2002) so
we limited comparison to stars with standard deviation of the {\it
I}-band magnitudes over the four years of OGLE-II observations smaller
than 0.05~mag, selecting in this way only the least variable objects. We
also limited the magnitude of stars to ${\it I} < 16.5$~mag.

\begin{figure}[p]

\centerline{\includegraphics[width=12.5cm, bb=40 50 510 560]{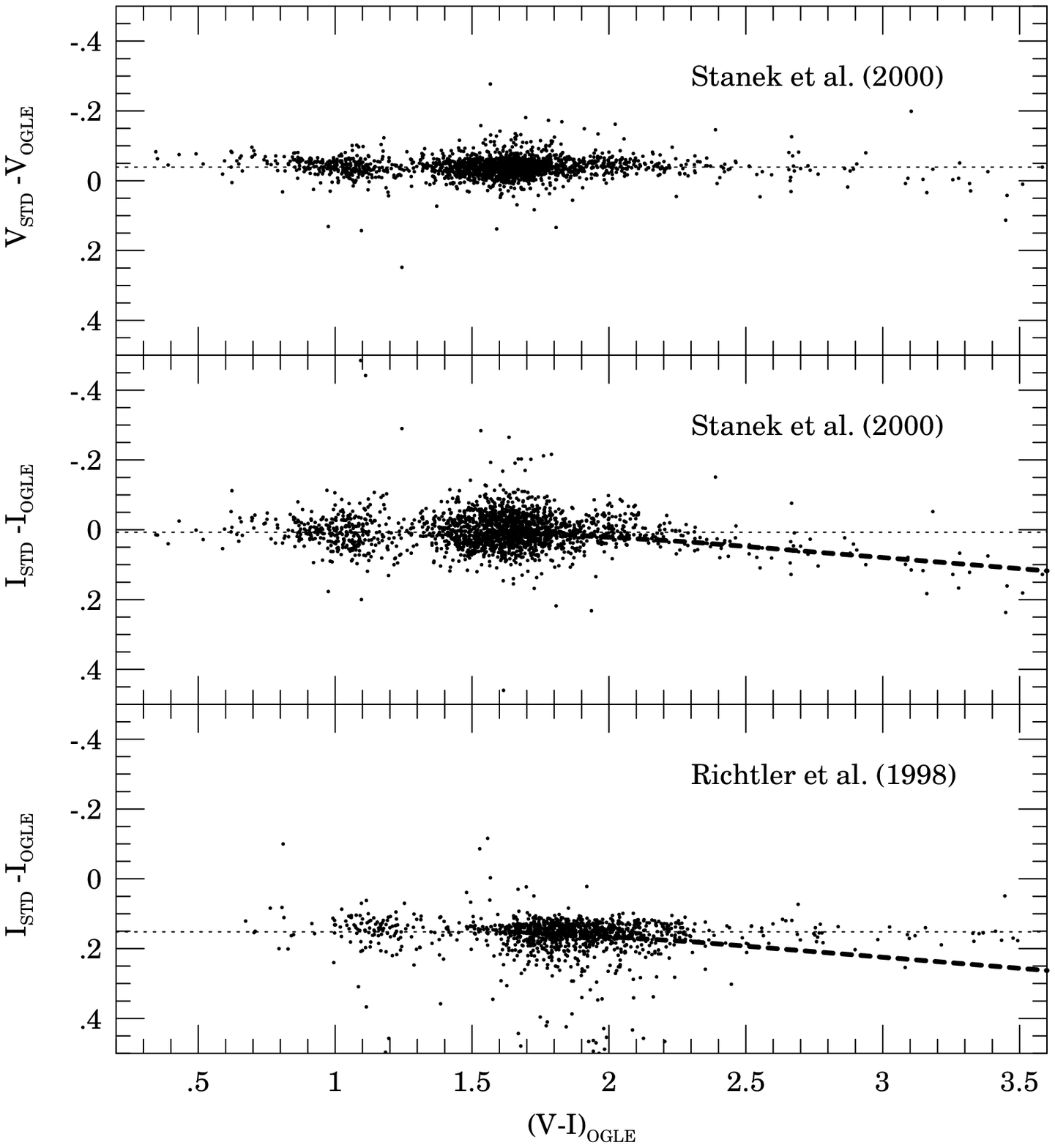}}
\FigCap{Comparison of the OGLE-II photometry with Stanek \etal (2000)
and Richtler \etal (1998). Dashed line in the middle and bottom panels
indicates differences between OGLE-II and standard Landolt {\it I}-band
filters as modeled by Udalski \etal (2002). Dotted line marks the mean
difference between the zero points.}
\end{figure}

Almost 2000 and 1170 stars  were matched between OGLE-II \vs Stanek
\etal (2000) and OGLE-II \vs Richtler \etal (1998) objects,
respectively. Fig.~7 presents the difference of magnitude as a function
of OGLE-II $V-I$ color. It is striking that only small number of red
stars ($V-I>2.2$~mag) passed the standard deviation limit confirming
that the vast majority of these objects are variable. The top panel
presents the comparison of {\it V} magnitudes between Stanek \etal
(2000) and OGLE-II. As noted by Udalski \etal (2002) one cannot expect
significant differences between the OGLE-II photometry and the standard
one in this band as the filters were similar. Indeed, the comparison in
the top panel of Fig.~7 indicates very good agreement -- only a small
shift of the zero point and probably a small slope of differences can be
noticed. The latter can be attributed to the possible small inaccuracies
in transformation from the instrumental system to the standard one by
Stanek \etal (2000) as their calibrations were based on observations
obtained on one night only.

Middle panel of Fig.~7 shows the comparison of the {\it I}-band
photometry of OGLE-II and Stanek \etal (2000). It should be stressed
that the observations by Stanek \etal (2002) were obtained with the
standard {\it I}-band filter. It is worth noting that the differences
between the OGLE-II photometry and the standard  one  behave exactly as
expected in Udalski \etal (2002). For the color range calibrated by
standards ($V-I<2$) they are constant while for redder stars the OGLE-II
magnitudes are systematically too bright. Dashed line in Fig.~7
indicates the difference of OGLE-II and standard magnitudes of RC stars
from this field ($V-I\approx 1.65$) as modeled in Udalski \etal (2002).
As one can see the agreement between the modeled and observed
differences is very good indicating that modeling presented in Udalski
\etal (2002) is accurate to 0.02--0.04~mag.

In the bottom panel of Fig.~7 we show comparison of the OGLE-II and
Richtler \etal (1998) photometry in the {\it I}-band. Again, except for
a shift of the zero point (see below), the agreement of photometries in
the range $V-I<2$ is very good. It is also very good for redder stars 
with the OGLE-II magnitudes being only slightly brighter
(0.02--0.03~mag) at $V-I$ of 3.5~mag. This behavior is somewhat
different than observed in the middle panel of Fig.~7 but Richtler's
\etal (1998) data were collected through the Gunn $i$ filter and later
transformed to the standard Landolt system  what can explain why the
differences in the bottom panel  are smaller than in the middle panel of
Fig.~7. We may conclude that our comparisons of OGLE-II photometry for
red stars with others show that the modeling presented in Udalski \etal
(2002) very reasonably estimates differences between the OGLE-II filter
and the standard one and, as expected, the modeling values are likely to
be an upper limit for the differences.  

Our comparisons also provide information on the accuracy of zero points
of the OGLE-II photometric maps. The mean differences between the
OGLE-II and Stanek \etal (2000) magnitudes for stars with $V-I<2.1$
(Stanek -- OGLE) are $-0.039, -0.042$ and $0.007$~mag for {\it V}, $V-I$
and {\it I}, respectively. All these values are within the estimated
errors of the OGLE-II zero points. On the other hand, the comparison
with photometry of Richtler \etal (1998) indicates systematic shift of
similar value in {\it V} and {\it I}  between both photometries: $0.133,
0.152$~mag for {\it V} and {\it I}, respectively, and $-0.014$~mag for
$V-I$ color (Richtler -- OGLE). However, the shift is of no concern:
Richtler \etal (1998) compared their photometry with the older one of
Ortolani, Bica and Barbuy (1992) and found shifts of similar values as
with OGLE-II magnitudes (Table~4 of Richtler \etal: $0.129$ and
$0.000$~mag for {\it V} and $V-I$, respectively, for stars brighter than
$V<18.5$~mag, what corresponds roughly to our limit of $I<16.5$~mag;
differences in the sense Richtler -- Ortolani). Very similar shifts
indicate that the OGLE-II photometry is almost identical with that of
Ortolani \etal (1992) and calibration of Richtler \etal (1998) data
suffers from quite large systematic error. It is worth noting here that
the OGLE-II calibrations were obtained from observations of standard
stars collected on tens of nights (Udalski \etal 2002) while the others
on single or two nights only, so they are certainly more prone to
systematic errors.

To compare the observed slopes of RC centroids with expected ones for
the standard interstellar extinction   we performed modeling similar as
in Udalski \etal (2002). We selected a model of atmosphere of a typical
bulge RC star from Kurucz (1992) archive (${\rm [Fe/H]}=-0.2$~dex, ${\rm
T}_{\rm eff}=4500$~K and $\log g = 2.00$). Such a star has $V-I\approx
1.0$.  Then we reddened it with the standard extinction law with ${\rm
R}_V=3.1$ and calculated the magnitudes for the standard and OGLE-II
filters. Unfortunately, the shape of the extinction curve longward of
the {\it V}-band has not yet been precisely determined. While it is
often assumed that its shape follows the CCM89 approximation, another
approach -- cubic spline interpolation between the ``anchor points'' --
was proposed by F99. The differences between both approximations are
large enough that we decided to perform our modeling for both cases
independently.  Lower plot in Figs.~2--6 corresponds to the modeling
with CCM89 extinction law, while the upper one (shifted by 1.2~mag along
the magnitude axis) to the F99 extinction law approximation.

Dashed lines in Figs.~2--6 show the position of the modeled RC star in
the CMD for the standard {\it I}-band filter. In other words they
indicate the standard reddening line in the observed regions. In Fig.~6
the line is not shown for CCM89 modeling, as it would overlap with the
RC fit line. Mean slopes of the standard reddening lines in the standard
{\it I}-band filter for both extinction law approximations are provided
in the third and fifth column of Table~1.

\MakeTable{|l|c|cc|cc|}{12.5cm}{Slope of the RC centroids and standard reddening line in
the OGLE-II CMDs}
{
\hline
\multicolumn{1}{|c|}{\dorule \uprule Region} & Observed Slope & 
\multicolumn{4}{|c|}{Slope of the Reddening Line (${\rm R}_V=3.1$)} \\
\cline{3-6}
& \dorule $\Delta I/\Delta (V-I)$ &
\multicolumn{2}{|c|}{\dorule \uprule CCM89} & \multicolumn{2}{|c|}{F99} \\
\cline{3-6}
& & \dorule \uprule Standard {\it I}& OGLE-II {\it I} & Standard {\it I} &
OGLE-II {\it I} \\
\hline
\uprule
(A) -- GB  & $0.94\pm0.01$ & 1.47 & 1.31 & 1.34 & 1.23\\
(B) -- GB  & $1.06\pm0.02$ & 1.47 & 1.32 & 1.34 & 1.23\\ 
(C) -- GB  & $1.16\pm0.03$ & 1.47 & 1.32 & 1.34 & 1.23\\ 
(D) -- GB  & $1.08\pm0.02$ & 1.46 & 1.34 & 1.33 & 1.24\\ 
\dorule
(E) -- LMC & $1.44\pm0.03$ & 1.46 & 1.39 & 1.30 & 1.25\\ 
\hline
}

Dotted lines in Figs.~2--6 mark the results of modeling for the OGLE-II
filter  transformed to the standard system as the real data were
(Udalski \etal 2002). They indicate the standard reddening line as it
should be observed through the OGLE-II filter. The mean slopes of the
standard reddening line in the OGLE-II filter are also listed in 
Table~1 (fourth and sixth column). As expected they are somewhat smaller
than for the standard {\it I}-band filter. The difference depends on
color, it is larger for redder $V-I$ colors and practically negligible
in the range where the OGLE data were calibrated by standards ($V-I<2$).
As we already mentioned, the differences presented in Figs.~2--6 are
very likely to be an upper limit.

Were the interstellar extinction close to the standard one, the RC
centers in Figs.~2--6 should be scattered around the dotted lines. This
is the case only in the LMC region for CCM89 approximation (Fig.~6 lower
plot). In the GB regions the slopes of the observed centers of RC are
generally  significantly smaller than the slope of the standard
reddening line in the OGLE-II filter. It is, however, worth noting that
the largest difference is found in the region (A), located closest to
the GC. In other regions the differences are smaller but do not seem to
depend on the distance from the GC. For instance, in the region (C),
located on the opposite side of the GB -- only about $3\arcd$ from the
GC -- the interstellar extinction seems to be the most similar to the
standard one.

It is evident that the different approximations of the extinction curve
give significantly different results for the slope of the reddening
line. For the CCM89 approximation the differences between the observed
RC slopes and  expected for the standard (${\rm R}_V=3.1$) extinction 
are larger than for the F99 approach. On the other hand the observed
slope in the LMC nicely agrees  with the slope of the standard reddening
line for CCM89 while it is too large for F99.

If the interstellar extinction toward the GB follows the standard
extinction law, then the smaller slopes of RC centroids  seen in
Figs.~2--5 can be explained by different (smaller) value of the ${\rm
R}_V$ parameter compared to the standard value of ${\rm R}_V=3.1$.
Therefore, we repeated our modeling of RC stars with different
parameters ${\rm R}_V$ of the standard extinction law, looking for
values of ${\rm R}_V$ for which the reddening line in the OGLE-II filter
would fit the observational data, \ie the solid and dotted lines in
Figs.~2--6 would overlap. Table~2 lists such values of ${\rm R}_V$ for
our regions as well as corresponding values of the ${\rm R}_{\it VI}$
ratio for the standard Landolt {\it I}-band. Values for both CCM89 and
F99 approximations of the extinction curve are provided.

\MakeTable{|l|c@{\hspace{29pt}}c|cc@{\hspace{-12pt}}|}{12.5cm}{Anomalous extinction in the OGLE-II fields}
{
\hline
\multicolumn{1}{|c|}{\uprule  \dorule Region} &
\multicolumn{4}{|c|}{\dorule\uprule Approximation of the extinction law} \\
\cline{2-5}
\dorule \uprule & \multicolumn{2}{|c|}{CCM89} & \multicolumn{2}{|c|}{F99} \\
\cline{2-5}
\dorule \uprule & ${\rm R}_V$ & ${\rm R}_{\it VI}$ & ${\rm R}_V$ & ${\rm R}_{\it VI}$ \\
\hline
\uprule
(A) -- GB  &  1.75 & 2.00 & 1.65 & 1.85 \\
(B) -- GB  &  2.15 & 2.15 & 2.10 & 2.05 \\ 
(C) -- GB  &  2.50 & 2.30 & 2.60 & 2.20 \\ 
(D) -- GB  &  2.15 & 2.15 & 2.20 & 2.05 \\ 
\dorule
(E) -- LMC &  3.30 & 2.50 & 4.60 & 2.55 \\ 
\hline
}                 

Table~2 shows that the observed flatter slopes of the centers of RC can
be explained by anomalous interstellar extinction toward the GB, \ie
extinction described by  smaller ${\rm R}_V$ or ${\rm R}_{\it VI}$ ratio
than the standard value. Resulting values of ${\rm R}_V$ in the GB
regions are roughly similar for both approximations of the extinction
curves -- CCM89 and F99, but corresponding values of ${\rm R}_{\it VI}$
are typically smaller by 0.1--0.15 for F99 approximation.  On the other
hand, the latter indicates highly non-standard extinction in the LMC.
According to the CCM89 approximation the properties of interstellar
extinction in the LMC are very similar to the standard ones. At this
stage it is premature to decide which approximation of the the red and
infrared extinction curve is better, but it seems that  additional more
extended studies similar to presented in this paper should shed the
light on the actual shape of the extinction curve at longer wavelengths.

As we showed, the anomalous interstellar extinction can explain the
observed location of RC centroids in the GB  fields. Are there other
possibilities? One could suspect that the mean brightness of the RC that
serves as a reference of light in our study is not constant. Indeed, it
is rather widely accepted that its mean {\it I}-band magnitude is
somewhat dependent on metallicity. However, the dependence is very weak
and well calibrated (Udalski 2000, Pietrzy{\'n}ski \etal 2003). To
change the slope from the standard to observed values in Figs.~2--5 the
metallicity distribution of RC stars in the GB would have to be highly
correlated with the reddening (larger reddening -- lower metallicity).
The metallicity gradients would have to be enormous in our small fields
to be responsible for the observed difference of slopes. This
possibility  seems to be extremely unlikely as the stars in the GB
should be well mixed and the population of RC stars very uniform,
especially in so small regions as analyzed in this paper. Moreover,
Ramirez \etal (2000) do not find any significant metallicity gradients
in the inner bulge including our regions.

Another possibility is that the mean RC in our subfields is located at
different distances -- again correlated with interstellar reddening
(larger reddening -- closer the mean RC). However, this situation also
seems to be unlikely. The range of photometry is always large enough
that we do not miss RC stars even in our most reddened subfields.  The
observed regions are small so it is hard to imagine large differences in
the mean distance of RC stars. Although Stanek \etal (1997) noted almost
0.4~mag difference in the brightness of RC in  regions located
symmetrically in longitude along the GC, and interpreted it as an effect
of geometry of the GB (assuming constant and standard interstellar
extinction), the spatial scale of such variation was about 10~degrees.
Therefore it is hard to expect geometrical effects larger than about
0.05~mag (or less if ${\rm  R}_{\it VI}$ is variable across the GB) in
our at most $1\arcd - 1\zdot\arcd3$ wide regions. This  is much too low
to explain the observed slopes of RC centers by geometry. For instance,
in the case of the region (A), the most reddened RC stars are  0.4--0.5
mag brighter than they would be in the  standard reddening case
(Fig.~2). So they would have to be by about 1.5~kpc closer to us compared
to the least reddened subfields in this region if the distance
differences (geometry) were to compensate the  difference of slopes.

Thus, the anomalous extinction seems to be the most natural explanation
of the observed slopes of RC in the GB.  The determined values of ${\rm
R}_V$ in our regions are much smaller than usually observed for stars
(CCM89), but it should be noted that similar low values were reported in
the literature for interstellar cirrus clouds (Szomoru and Guhathakurta
1999). One of such regions is located only $10\arcd$--$14\arcd$ from our
fields so it would not be surprising if large parts of the GB regions
were embedded and obscured by similar cirrus clouds. Our results
indicate that ${\rm R}_V$ and the ${\rm R}_{\it VI}$ ratio are variable
across the GB and variations are significant. This is very unfortunate
as it indicates that proper dereddening of a particular field in the GB
might be difficult without prior determination of ${\rm R}_V$ in a given
line-of-sight.

It should be noted that the models of the Galactic bar based on modeling
of RC stars (Stanek \etal 1997) might be somewhat affected by our
finding, because Stanek \etal (1997) assumed constant and standard ${\rm
R}_{\it VI}$. While our analysis indicate variable and non-standard 
interstellar extinction in the GB, the range of variation of ${\rm 
R}_{\it VI}$ (Table~2) is much too small to fully account for effects
presented by Stanek \etal (1997). Therefore, we suspect that only some
fine tuning to the GB models might be necessary, when accurate ${\rm 
R}_{\it VI}$ in the lines-of-sight presented in Stanek \etal (1997) are
known.

It is also worth noting that our results nicely confirm predictions of
Popowski (2000). Although Stanek \etal (2000) suggested that the
difference between the mean dereddened (with the extinction map of
Stanek 1996 based on standard ${\rm  R}_{\it VI}$=2.5) $V-I$ color of
the GB and local (Hipparcos)  RC stars is small and within uncertainties
of photometry, we note that the GB stars are still about 0.06~mag redder
according to their data. OGLE-II photometry of exactly the same region
indicates, however, that the difference is larger, \ie 0.09~mag -- in
good agreement with the difference of the $V-I$ color zero point between
photometries found above. This value is also very consistent with the
result of  Paczy{\'n}ski \etal (1999) in another region of GB. As we
believe that the zero points of the OGLE-II photometry are sound (see
above) we may conclude that the difference between the mean dereddened
in that way $V-I$ colors of the GB and local RC stars is real at the
0.06--0.09~mag level. 

According to Table~2, ${\rm  R}_{\it VI}$ is equal to 2.1--2.2 for the
Baade's window (region (D)) for both approximations of the extinction
law. If such a  value of ${\rm R}_{\it VI}$ is used to deredden
photometry of the GB stars instead of the standard one (used in the map
of Stanek 1996) then  the difference in the $V-I$ colors of RC and RR
Lyr stars from the GB and local population (Paczy{\'n}ski and Stanek
1998, Stutz \etal 1999) diminishes by additional 0.08--0.10~mag to very
small values well explained by uncertainties of photometry (Popowski
2000).

Summarizing, the OGLE-II photometry seems to indicate that the
interstellar extinction toward the GB is anomalous. However, additional
observations are necessary for further verification and confirmation of
this finding. Photometry in larger number of preferably narrower bands
should provide more information on the shape of the red and infrared
extinction curve. Alternatively, spectrophotometry of large sample or RC
stars from the GB should probably allow to disentangle effects of
extinction, metallicity and other population effects in RC stars spectra
and determine the extinction properties in much wider range, likely with
better accuracy than it is possible with photometry alone. A list of
possible RC stars suitable for such a project can be prepared using the
OGLE-II GB maps (Udalski \etal 2002).

\Acknow{We would like to thank Dr.\ B.~Draine for very important
comments and pointing our attention to F99 approximation of the
extinction law and Drs.\ B.~Paczy\'nski, K.Z.\ Stanek and
M.~Szyma{\'n}ski for many helpful remarks. We also thank the anonymous
referee for constructive criticism leading to significant improvement of
the paper.  The paper was partly supported by the Polish KBN BST grant
to Warsaw University Observatory. Partial support  for the OGLE project
was provided with the NSF grants AST-9820314 and AST-0204908 and NASA
grant NAG5-12212 to  B.~Paczy\'nski.}

\end{document}